\documentclass[apjl]{emulateapj-rtx4}
\usepackage{graphicx}
\usepackage{natbib}

\begin{document}     

\title{Metallicity gradients through disk instability: a simple model \\ for the Milky Way's boxy bulge }

\shorttitle{Vertical metallicity gradients}

\author{Inma Martinez-Valpuesta\altaffilmark{1} and Ortwin Gerhard\altaffilmark{1}}
\affil{Max-Planck-Institut f\"ur Extraterrestrische Physik, Giessenbachstrasse, 85748 Garching, Germany}
\email{imv @ mpe.mpg.de, gerhard @ mpe.mpg.de}
\shortauthors{Inma~Martinez-Valpuesta \& Ortwin~Gerhard}

\label{firstpage}

\begin{abstract}
  Observations show a clear vertical metallicity gradient in the
  Galactic bulge, which is often taken as a signature of dissipative
  processes in the formation of a classical bulge.  Various evidence
  shows, however, that the Milky Way is a barred galaxy with a boxy
  bulge representing the inner three-dimensional part of the bar. Here
  we show with a secular evolution N-body model that a boxy bulge
  formed through bar and buckling instabilities can show vertical
  metallicity gradients similar to the observed gradient, if the
  initial axisymmetric disk had a comparable radial metallicity gradient.
  In this framework the range of metallicities in bulge fields
  constrains the chemical structure of the Galactic disk at early
  times, before bar formation. Our secular evolution model was
  previously shown to reproduce inner Galaxy star counts and we show
  here that it also has cylindrical rotation. We use it to predict a
  full mean metallicity map across the Galactic bulge from a simple
  metallicity model for the initial disk. This map shows a general
  outward gradient on the sky as well as longitudinal perspective
  asymmetries. We also briefly comment on interpreting metallicity
  gradient observations in external boxy bulges.
\end{abstract}


\section{Introduction}
\label{sec:intro}

About half of edge-on disk galaxies contain boxy or peanut-shaped
bulges \citep[BPBs,][]{Luetticke+00}.  Their photometric and kinematic
properties are consistent with predictions from disk galaxy
simulations in which a BPB formed through bar and buckling
instabilities \citep{Bureau+Freeman99, Debattista+04,
  Bureau+Athanassoula05}. The Galactic bulge shows many
characteristics of a BPB formed in this way:
a boxy shape in projection \citep{Dwek+95, Skrutskie+06}; 
a triaxial density distribution
\citep{Binney+97,Lopez-Corredoira+05} with a
dense inner, rounder component \citep[][hereafter
GMV12]{Gonzalez+12, Gerhard+Martinez-Valpuesta12};
an X-shaped structure \citep{McWilliam+Zoccali10,Nataf+10,
  Ness+12a, Li+Shen12};
cylindrical rotation \citep{Howard+09, Shen+10};
and a transition to an outer planar bar \citep[hereafter
MVG11]{Martinez-Valpuesta+Gerhard11},  seen in star count
observations as 'long bar' \citep[][]{Benjamin+05,
  Cabrera-Lavers+07}.

The Galactic bulge is also known to consist of predominantly old
stars \citep{Zoccali+03, Clarkson+08} with a broad,
asymmetric metallicity distribution function (MDF)
\citep{Rich90,Ibata+Gilmore95b,Zoccali+03}.  The data clearly show a
vertical metallicity gradient, such that the more metal-rich part of
the MDF thins out towards high latitudes \citep[$b <
-4\degr$,][]{Minniti+95,Zoccali+08,Gonzalez+11}. At latitudes
lower than Baade's window ($-4\degr<b<0\degr$) no vertical gradient is
seen \citep{Rich+07, Rich+12}.  These properties are not obviously 
reconciled with a disk origin of the bulge. Because bars
tend to mix stars from different radii and so wash out preexisting
metallicity gradients \citep{Friedli+94}, the observed metallicity
gradient in the Galactic bulge has long been taken as a signature for
a classical bulge in the Milky Way. Indeed, \citet{Bekki+Tsujimoto11}
found that they could not reproduce the observed gradient with a
simulated boxy bulge which had evolved from a pure disk with an
initial metallicity gradient.  In order to obtain the lower mean
metallicities at high $b$, they needed to include an additional
metal-poor thick disk in their initial model.

Here we show with the help of a suitable boxy bulge/bar simulation,
that contrary to widespread belief boxy bulges can have metallicity
gradients similar to those observed in our Galaxy and in other boxy
bulges, if they evolve from disks with similarly steep radial metallicity
gradients.

\section{N-body model for the Galactic bar and boxy bulge}
\label{sec:model}

The simulation used in this work is that already analyzed in MVG11 and
GMV12, with similar characteristics as described in
\citet{Martinez-Valpuesta+06}. The model evolved from an exponential
disk, with $Q=1.5$, initial radial scale length of $h=1.29$~kpc and
vertical scale height of $h_z=0.225$~kpc, embedded in a dark matter
halo. Following bar and buckling instabilities it developed a
prominent boxy bulge which is already relaxed at $\sim1.5$~Gyr.  As in
GMV12 we consider the simulated galaxy at time $T_b\simeq1.9$~Gyr when
the bar has resumed its growth through further angular momentum
transfer to the halo, and we vertically symmetrize the particle
distribution to optimize the resolution.

\begin{figure}
\begin{center}
\includegraphics[scale=0.4,angle=-90.]{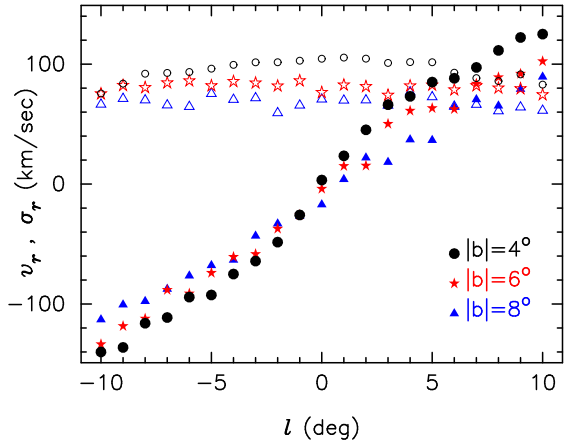}
\caption{Mean radial velocity and velocity dispersion versus longitude
  for the model's boxy bulge, as seen from the Sun for different
  latitudes $|b|=4\degr,6\degr,8\degr$. Note how the dispersion
  decreases with latitude, and that in this projection the rotation is
  only approximately cylindrical; i.e., the slope of the velocity
  curve increases slightly towards the plane, as also seen in the
  Galactic bulge data.}
\label{fig:kinematics}
\end{center}
\end{figure}

We showed previously that this model reproduces different Milky Way
star count observations, both near the Galactic plane for the long bar
out to $l\sim27\degr$ (MVG11), and in the inner boxy bulge for
$-10\degr\leq l \leq 10\degr$ and $b=\pm1\degr$ (GMV12) as well as for
$b=-5\degr$, where the predictions of the model were subsequently
confirmed by \citet{Gonzalez+12}. Here we show that the model's boxy
bulge at that time also shows approximate cylindrical rotation, with
the slope of the mean velocity curve along $l$ increasing slightly
towards lower latitudes; see Fig.~\ref{fig:kinematics}. This is
similar to the BRAVA kinematic data \citep{Howard+09} and the
simulation by \citet{Shen+10}. The cylindrical rotation in boxy bulges
is clearest in edge-on view, as shown in previous simulations
\citep[e.g.,][]{Combes+90,Athanassoula+Misiriotis02}; it is weakened
by the nearly end-on orientation of the Milky Way bar and the
perspective effects.  As in our previous work (MVG11, GMV12), we here
take the solar galactocentric distance $D_{\sun}=8$~kpc, scale the bar
length to $4.5$~kpc, and use a bar orientation of $\alpha=25\degr$
with respect to the Sun-Galactic center line.

We study the formation of bulge metallicity gradients in the unstable
disk -- boxy bulge scenario with a simple illustrative model. We
assume that the initial exponential disk follows a specified radial
metallicity gradient, which we imagine is set up during disk build-up
prior to bar formation. We assign to each particle a metallicity
depending on its initial radius \citep[e.g.,][]{Friedli+94}, according
to $[M/H](R)=[M/H]_0+\alpha_{R}\times R/{\rm kpc}$. After some
exploration we choose $[M/H]_0=0.6$ and $\alpha_{R}=-0.4$. Unlike
\citet{Bekki+Tsujimoto11}, we do not link these metallicities to the
present-day Galactic disk near the Sun, because the buckling
instability in the Milky Way must have happened long ago when the
outer disk may have been only incompletely assembled.

\section{Results}

\subsection{Vertical metallicity gradient in the Milky Way bulge}
\label{sec:vgrad}

\begin{figure}
\begin{center}
\includegraphics[scale=0.35,angle=-90.]{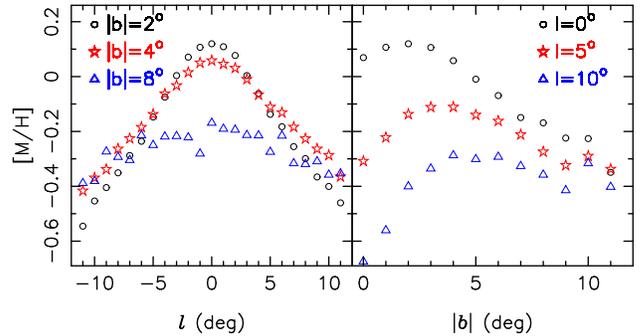}
\caption{Longitudinal and vertical metallicity profiles of the model
  as seen from the Sun. Left: with $l$ for $\vert
  b\vert=2\degr,4\degr,8\degr$; the gradient along $l$ decreases
  towards larger distances from the Galactic plane. Right: with $\vert
  b\vert$ for $l=0\degr,5\degr,10\degr$; the vertical gradient
  decreases with distance from the minor axis.}
\label{fig:rv_grad}
\end{center}
\end{figure}

The bar and buckling instabilities leading to the formation of the
boxy bulge strongly change the particle orbits. Thereby the initial
disk metallicity distribution is mapped into the bulge accordingly.
Fig.~\ref{fig:rv_grad} shows longitudinal and vertical metallicity
profiles for the bulge at $T_b$. The vertical gradient for $3\degr <
\vert b \vert <7\degr$ along the minor axis is
$\alpha_{b}=-0.064$~dex/deg, compared to the observed slope in the
Milky Way bulge $\alpha_{MW}=-0.075$~dex/deg from \citet{Zoccali+08}
and $\alpha_{b}=-0.06$~dex/deg from \citet{Gonzalez+11}.  Near the
Galactic plane, $\vert b \vert <3\degr$, the vertical gradient becomes
flatter and even positive.  The positive gradient is due to the
contamination from foreground/background disk particles.  The flat
part comes from the mixing during the buckling instability in the
inner regions that evolve into the central near-spheroidal component
(GMV12), and also from the greater fraction of low-latitude outer bar
particles.  The absence of a clear vertical gradient near the Galactic
plane is consistent with recent results from \citet{Rich+07,
  Rich+12}. However, their measured mean metallicities for M-giants
([Fe/H]$\simeq-0.2$) are slightly lower in Baade's window than those
for K-giants \citep{Zoccali+08} and for our simple model.

\begin{figure*}
\begin{center}
\includegraphics[scale=0.48,angle=-90.]{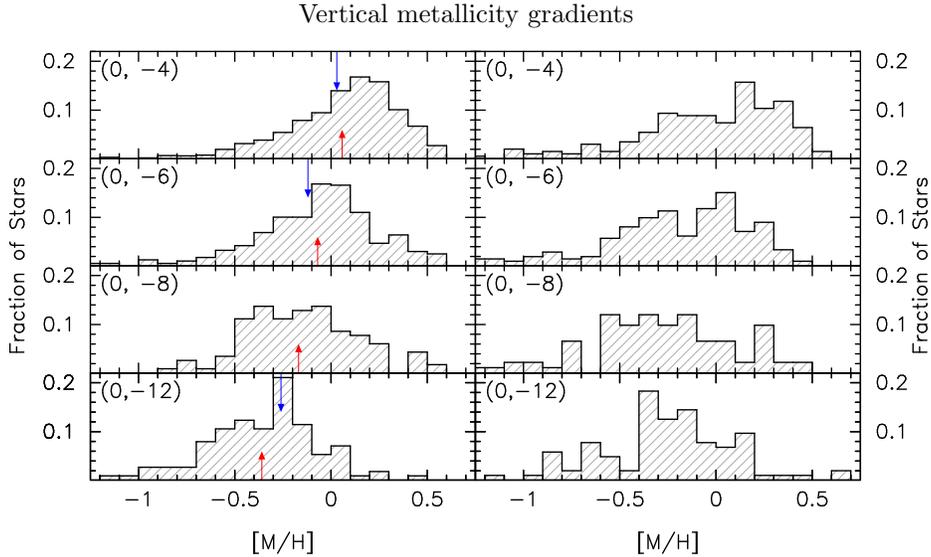}
\caption{Metallicity histograms (MDFs) for model particles in $4$
  fields along the minor axis of the boxy bulge ({\it left
    panels}). The mean value for each MDF is given by the red arrow
  pointing up from the bottom. The corresponding mean value for the
  data of \citet{Zoccali+08} is indicated by the blue arrow from the
  top. The histograms in the {\it right panels} are based on data from
  \citet{Zoccali+08} for $b=-4\degr, -6\degr, -12\degr$, and from
  \citet{Johnson+11} for $b=-8\degr$.}
\label{fig:vgrad}
\end{center}
\end{figure*}

Fig.~\ref{fig:vgrad} shows the resulting MDFs in several minor axis
fields in the boxy bulge at $T_b=1.9$ Gyr. Note the clear gradient of
mean metallicity with latitude, and that similarly to the observed
histograms, it arises mostly from the decreasing fraction of higher
metallicity stars away from the Galactic plane. The maximum
metallicity in all histograms is $\simeq0.6$, the central value in the
initial disk, and the lowest value is $\simeq-1.2$.  2 Gyr later, at
$T_b=4 $ Gyr, the model metallicity gradient is nearly identical,
while the mean metallicities in the minor axis fields have slightly
decreased, due to the capture of lower-metallicity disk stars by the
slowly growing bar.  Had we assumed a shallower radial gradient in the
initial disk, also the final vertical gradient in the bulge would be
smaller.

\subsection{Full metallicity map for the boxy bulge}
\label{sec:map}

\begin{figure}
\begin{center}
\includegraphics[scale=0.57,angle=-90.]{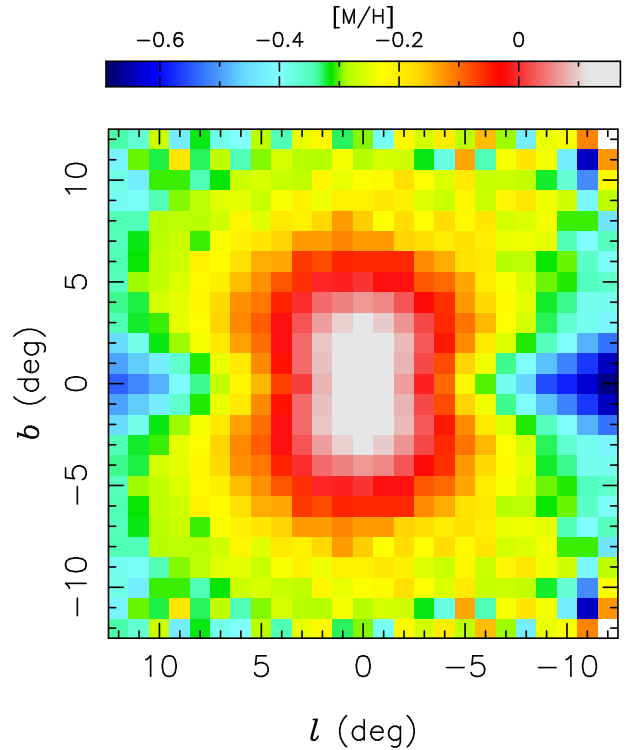}
\caption{Metallicity map of the model bulge and bar in galactic
  coordinates.  Foreground and background disk particles with distances
  $<4$~kpc and $>12$~kpc from the solar position are excluded. The
  colour in each square corresponds to the average metallicity in a
  cone with radius $0.5\degr$ centered at this position.  }
\label{fig:map_grad}
\end{center}
\end{figure}

So far we have shown that a radial metallicity gradient in the
unstable initial disk can survive through the bar and buckling
instabilities, and generate, for suitable parameters, a vertical
metallicity gradient similar to that observed in the Milky Way bulge.
For comparison with upcoming survey results (e.g., VVV, ARGOS,
Gaia-ESO), we now provide a full metallicity map, which gives a
different view of the predicted metallicities for this scenario.
Differently from Fig.~\ref{fig:vgrad}, we exclude
foreground/background disk stars with a distance cut, using only
particles with $4<D<12$~kpc from the position of the Sun.  The assumed
initial disk metallicity distribution and simulated galaxy snapshot
are as before.  Fig.~\ref{fig:map_grad} shows the resulting average
bulge metallicity map over the area on the sky extended by the
Galactic bulge.  It has several noteworthy properties:

(i) The approximate outline of the boxy bulge can be seen together
with low-metallicity indentations in the Galactic plane. The latter
are due to foregroud/background stars in the planar bar, which
in these positions dominate the bulge stars.

(ii) A metallicity gradient is present in all directions on the sky,
both vertically and radially. This is expected from the binding energy
argument of Sect.~\ref{sec:mechanism}. However, in the central few
degrees the gradients become shallower.

(iii) Iso-metallicity contours are more elongated vertically than
horizontally, whereas the surface density contours are flattened
to the plane (MVG11).

(iv) The asymmetry between $l>0$ and $l<0$ at intermediate latitudes
is due to perspective effects and signifies the bar origin. In
Fig.~\ref{fig:map_grad} high metallicities extend to larger $\vert
l\vert$ on the $l>0$ side at $\vert b\vert\simeq 5\degr$, but for
$\vert b\vert \simeq 10\degr$ they extend to larger $\vert l\vert$ on
the $l<0$ side.

This map has some remarkable similarities with that being constructed
from the near-infrared photometric VISTA VVV survey \citep{Gonzalez+13}. 
For now we can compare quantitatively with
a few more measured values in off-axis fields.  Details will of course
depend on the parametrisation of our simple model.  In the
longitudinal direction the gradient along $b=2\degr$ is
$\alpha_{l}=-0.05$~dex/deg (see Fig.~\ref{fig:rv_grad}).
\citet{Minniti+95} found mean metallicities [Fe/H]$=-0.3$ in two
fields at $(l,b)=(8\degr,-7\degr)$ and $(10\degr,-7\degr$) for bulge
stars selected with [Fe/H]$>-1$. The model values in these fields are
-0.30 and -0.28.

\subsection{Origin of the vertical gradient}
\label{sec:mechanism}

\begin{figure}
\begin{center}
\includegraphics[scale=0.6,angle=-90.]{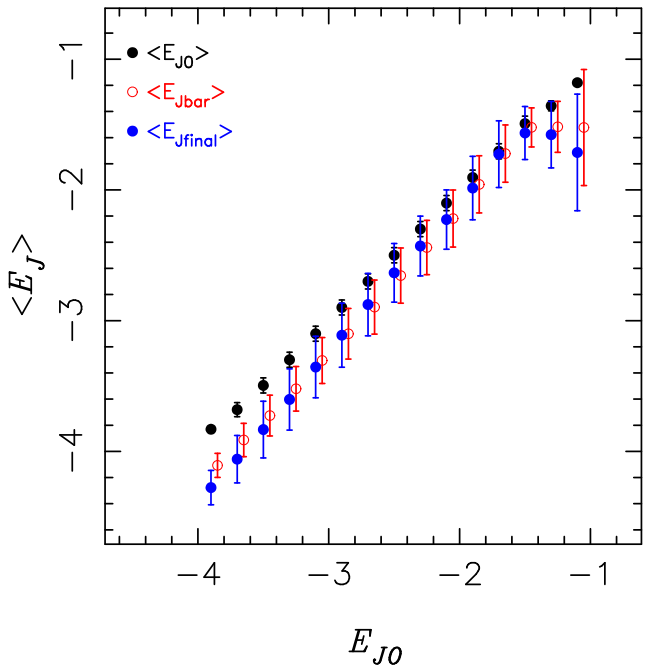}
\caption{Change of Jacobi energy between different times during the
  evolution of the model, evaluated in a rotating frame for pattern
  speed $\Omega_p(T_b)= 40$~km/s/kpc.  For each bin in initial Jacobi
  energy $E_{\rm J0}$, the mean value of Jacobi energy at time $t$ is
  plotted with error bars denoting the standard deviation of the
  distribution of $E_{\rm J}(t)$ in the bin.  Black points:
  $E_{J0}(t=0)$ versus $E_{J0}$ (initial scatter in each bin).  Blue
  points: $E_{J{\rm final}}(t=T_b)$ versus $E_{J0}$ (final boxy bulge
  compared to initial disk). Red open circles (slightly displaced
  horizontally for clarity): $E_{J{\rm bar}}(t=T_{\rm bar})$ versus $E_{J0}$
  (after full bar growth but prior to buckling instability, compared
  to initial disk). }
\label{fig:energy}
\end{center}
\end{figure}

Clearly, while the bar and buckling instabilities scramble the orbits
of disk stars, they do not do so enough to completely erase the
preexisting metallicity gradient. High-metallicity stars tightly bound
to the Galactic center initially remain more tightly bound in the
final bulge, and initially more loosely bound stars with lower
metallicities end up at larger final radii, on average. This can be
quantified by considering the change in Jacobi energy
\begin{equation}
E_{\rm J} = {1\over 2} v^2 + \Phi(R,\phi,z) - {1\over 2} \Omega_p^2 R^2
\end{equation}
in the rotating frame of the boxy bulge and bar during the
evolution. Here we use standard cylindrical coordinates, $v$ is total
velocity, $\Phi$ is the gravitational potential, and $\Omega_p$ is the
pattern speed at $T_b=1.9$ Gyr. Fig.~\ref{fig:energy} shows that
particles are scattered in Jacobi energy by the two instabilities, but
only over a fraction of the available total range in $E_{\rm
  J}$. I.e., they retain some memory of their initial values.  It has
also been shown that most bar particles inside the vertical inner
Lindblad resonance (VILR) mix during the buckling instability, but
stay in the inner bar regions, while most bar particles around VILR
are scattered to orbits which can visit larger heights
\citep{Pfenniger+Friedli91,Martinez-Valpuesta+Shlosman04}.

From these arguments follows that the number of high-metallicity
particles in the final boxy bulge will decrease with height above the
plane where $E_{\rm J}$ becomes less negative. In the
histograms of Fig.~\ref{fig:vgrad}, therefore, the number of stars on
the metal-rich part of the distribution decreases with latitude.  The
maximum metallicity in all histograms, but that at $b=-12\degr$, is
still given by the assumed maximum metallicity at the center of the
initial disk, here $+0.6$, because a small fraction of the tightly
bound particles is scattered up even to $b=8\degr$.

The low-metallicity tail, on the other hand, is more prominent at high
latitudes, because of the larger fraction of particles coming from the
outer parts of the bar. Note that within the assumed model, a lower
limit to the metallicity in the bulge fields is expected, which is set
by the outer boundary of the part of the disk which participates in
the instability. From the histograms in Fig.~\ref{fig:vgrad}, this is
$\simeq -1.2$, which is due to particles in the initial disk at $R=4.5
{\rm kpc}$, just inside the corotation radius of the bar before
buckling, $4.7$ kpc. 

These properties of the model histograms are surprisingly similar to
those of the Galactic bulge MDFs discussed in the literature.  If our
scenario were the full explanation for the metallicity structure of
the Galactic bulge, the observed range in the metallicity histograms
could be used directly to estimate the central metallicity and the
radial gradient in the inner Galactic disk prior to bar formation and
buckling. The observational histograms, however, also contain
  metal-poor stellar halo and thick disk stars, and possibly
  metal-rich disk stars formed after the buckling, which must be taken
  into account in this argument.  Finally we note that a radial
  metallicity gradient in the precursor disk could have been
  accompanied by a stellar age gradient.  In the context of the
  present model, depending on the time of buckling, this could lead to
  a measurable vertical age gradient in the boxy bulge which would
  provide an independent test of the model.  

\subsection{Boxy bulges in other galaxies}

Vertical gradients have so far been published only for a small number
of galaxies with boxy bulges. \citet{Falcon-Barroso+04, Jablonka+07}
found a vertical metallicity gradient in NGC 7332 comparable to that
in the Milky Way. \citet{Williams+11} determined vertical gradients in
two other boxy bulges, NGC 1381 and NGC 3390. They integrated along
slits parallel to the major axis to increase signal-to-noise, and
obtained a single bulge measurement at several $z$.  A similar
averaging over $l\in[-5\degr,5\degr]$ in our model reduces the
measured gradient from $\alpha_z=-0.46$~dex/kpc to
$\alpha_z=-0.33$~dex/kpc, because radial and vertical gradients
are comparable.  Still, in NGC~1381 the observations show a strong
gradient of averaged $\alpha_z=-0.27$~dex/kpc, while only a weak
gradient is seen in NGC~3390, $\alpha_z=-0.1$~dex/kpc
($z\in[5,10]$~arcsec).  \citet{Perez+Sanchez-Blazquez11} found
negative metallicity gradients along the {\sl major axis} for the
majority of bulges in barred galaxies, while \citet{Williams+12} found
a range of major axis metallicity gradients in boxy bulges, from
negative to positive, with shallower sample average than for unbarred
early-type galaxies. In summary, the limited data suggest a range of
metallicity gradients in boxy bulges, with the metallicity gradient in
the Milky Way on the steep side. In the framework discussed here,
this suggests a range of initial disk metallicity gradients before
buckling in these galaxies, perhaps depending on the time of bulge
formation.

\subsection{Comparison with previous simulations}

The results presented in the previous subsections are consistent with
early N-body simulation work by \citet{Friedli98}. He found that
pre-existing {\sl vertical} abundance gradients in the disk were
quickly flattened both in the bar and disk regions but not entirely
erased. Their models also included a shallow, outward {\sl radial}
gradient, which became much flatter in the disk region due to the
mixing induced by the bar, but was approximately preserved in the bar
region.  \citet{Friedli98} also stated that when no initial vertical
gradient was present, a negative vertical gradient appeared at
$R=0$. However, in his model the initial gradient was
$\alpha_{R}=-0.1$~dex/kpc, while the model analyzed here has initial
$\alpha_{R}=-0.4$~dex/kpc, for similar bar size. In our model the
final gradient along the bar is $\alpha_{R}=-0.26$~dex/kpc and the
final vertical gradient is $\alpha_z=-0.46$~dex/kpc (for
$b\in[-3\degr,-7\degr]$).  Consistent with \citet{Friedli98} we also
find that at larger $l$ the vertical gradient is shallower
(Fig~\ref{fig:rv_grad}, {\it right}).

Recent N-body numerical simulations by \citet{Bekki+Tsujimoto11} for a
pure exponential disk initial model (PDS in their nomenclature) showed
only a very shallow vertical gradient in the final boxy bulge.  The
difference to our results can be traced to the different initial disk
metallicity profile.  While their inner disk parameters are similar to
ours, due to linking the profile to the present metallicity near the
Sun, their profile becomes quite shallow beyond $\simeq 0.25$ times
the eventual final bar length.  Thus the stars in the upper bulge
which come from the outer parts of the bar are quite metal-rich.
The difference, and the reason for the steep vertical gradient
in the bulge of our model is its steep initial radial gradient in the
disk.

Finally we note that if the secular evolution proceeds slowly, a
radial gradient in the disk before the instability may be (partially)
erased by migration processes. Thus a steep final bulge gradient is
favoured by rapid secular evolution such as in the model investigated
here.

\section{Conclusions}

We can summarize our conclusions as follows:

(i) The vertical metallicity gradient observed in the Galactic bulge
can be reproduced with a secularly evolved barred galaxy model in
which a boxy bulge formed after a buckling instability. In itself, a
vertical gradient is therefore not a strong argument for the existence
of a classical bulge in the Milky Way.

(ii) Mixing during the bar and buckling instabilities is incomplete,
and therefore radial metallicity gradients in the initial disk can
transform into gradients in the boxy bulge.

(iii) In this framework, the range of bulge star metallicities at
various latitudes constrains the radial gradient in the precursor
disk.

(iv) The full bulge metallicity map shows an overall radial gradient
on the sky as well as longitudinal perspective asymmetries. 
Iso-metallicity contours are elongated vertically.

(v) Depending on when the bulge formed and on the properties of the
precursor disk at that time, boxy bulges may or may not show
metallicity gradients.

The success of our simple model in explaining Milky Way observations
suggests that its underlying idea has some merit and should be pursued
further.  Future work also needs to consider chemical evolution and
possible vertical metallicity gradients in the initial disk, star
formation induced by the bar, the possible late build-up of a metal
rich inner disk, and the contribution of halo stars and other components
\citep{Ness+12b} to the bulge MDFs.

Another important next step for understanding the origin of the Milky
Way bulge is to investigate the correlations between metallicities and
kinematics in this model and in generalisations of it. Comparison with
similar observations \citep{Babusiaux+10, Ness+12b} is likely to shed
light on the possible multi-component nature of the Galactic
bulge. The results will be important also for interpreting the data
for bulges in external galaxies.

\acknowledgments

We thank Ken Freeman and Mike Williams for discussions on metallicity
gradients and Oscar Gonzalez for showing us the VVV metallicity map
before publication.

\bibliographystyle{apj}

\label{lastpage}
\end{document}